\documentclass[twocolumn,showpacs,preprintnumbers,amsmath,amssymb,superscriptaddress,prc,nofootinbib]{revtex4}

\usepackage{scalerel}

\usepackage{graphicx}
\usepackage{dcolumn}
\usepackage{bm}
\usepackage{amsmath,amssymb}
\usepackage[dvipsnames]{xcolor}
\usepackage{lineno}

\makeatletter
\newsavebox{\@brx}
\newcommand{\llangle}[1][]{\savebox{\@brx}{\(\m@th{#1\langle}\)}%
  \mathopen{\copy\@brx\kern-0.5\wd\@brx\usebox{\@brx}}}
\newcommand{\rrangle}[1][]{\savebox{\@brx}{\(\m@th{#1\rangle}\)}%
  \mathclose{\copy\@brx\kern-0.5\wd\@brx\usebox{\@brx}}}
\makeatother

\begin{document}


\title{Statistical uncertainties of the $v_{n}\{2k\}$ harmonics from Q-cumulants}

\author{L. Nadderd}
\affiliation{Department of physics, ``VIN\v{C}A'' Institute of Nuclear Science - National  Institute of the Republic of Serbia, University of Belgrade, Belgrade, Serbia.}
\affiliation{Zhejiang  -  Serbia  Strong  coupling  Physics  Research  Laboratory, Huzhou  University,  Huzhou,  Zhejiang  313000,  P.  R.  China.}

\author{J. Milosevic}
\email[Corresponding author:]{Jovan.Milosevic@cern.ch}
\affiliation{Department of physics, ``VIN\v{C}A'' Institute of Nuclear Science - National  Institute of the Republic of Serbia, University of Belgrade, Belgrade, Serbia.}
\affiliation{Zhejiang  -  Serbia  Strong  coupling  Physics  Research  Laboratory, Huzhou  University,  Huzhou,  Zhejiang  313000,  P.  R.  China.}

\author{F. Wang}
\affiliation{Zhejiang  -  Serbia  Strong  coupling  Physics  Research  Laboratory, Huzhou  University,  Huzhou,  Zhejiang  313000,  P.  R.  China.}
\affiliation{College of Science, Huzhou University, Huzhou, Zhejiang 313000, P. R. China.}
\affiliation{Department of Physics and Astronomy, Purdue University, Indiana 47907, USA.}

\date{\today}

\begin{abstract}
  Analytic formulas to calculate statistical uncertainties of $v_{n}\{2k\}$ harmonics extracted from the Q-cumulants are presented. The Q-cumulants are multivariate polynomial functions of the weighted means of $2m$-particle azimuthal correlations, $\llangle 2m \rrangle$. Variances and covariances of the $\llangle 2m \rrangle$ are included in the analytic formulas of the uncertainties that can be calculated simultaneously with the calculations of the $v_{n}\{2k\}$ harmonics. The calculations are performed using a simple toy model which roughly simulates elliptic flow azimuthal anisotropy with magnitudes around 0.05. The results are compared with the results obtained by the many data sub-sets, and by the bootstrapping method. The first one is a common way of estimation of the statistical uncertainties of the $v_{n}\{2k\}$ harmonics in a real experiment. In order to increase precision in the measurement of the $v_{n}\{2k\}$ harmonics, a large number of 15000 sub-sets and the same number of the re-sampling in the bootstrap method is used. Unlike the other ways of the analytic calculation of the $v_{n}\{2k\}$ statistical uncertainties, our proposal that includes the use of squared weights in the calculation of both the variances and covariances, gives the best agreement with the results obtained from the sub-sets and bootstrap method. Additionally, a recurrence equation between Q-cumulants of any order is also presented.
\end{abstract}



\maketitle

\section{Introduction}

\label{intro}
In high energy nucleus-nucleus collisions, the Quark-Gluon-Plasma (QGP)~\cite{Arsene:2004fa,Back:2004je,Adams:2005dq,Adcox:2004mh}, a state of matter formed from a large number of deconfined quarks and gluons has been created and studied. The collective anisotropic expansion of the QGP is one of important phenomena used to such studies~\cite{Aamodt:2010pa,ALICE:2011ab,Abelev:2014pua,Adam:2016izf,ATLAS:2011ah,ATLAS:2012at,Aad:2013xma,Chatrchyan:2012wg,Chatrchyan:2012ta,Chatrchyan:2013kba,CMS:2013bza,Khachatryan:2015oea}. Many methods have been developed~\cite{Ollitrault:1993ba,Voloshin:1994mz,Poskanzer:1998yz}. The first historical references~\cite{Borghini:2001vi,Borghini:2000sa} in which the method of cumulants have been introduced in flow analyses enabled suppression of the short-range correlations arising from jets and resonance decays and thus revealed the collective nature of the QGP. An improved method of cumulants is proposed in~\cite{Bilandzic:2010jr}. This method is referred to as Q-cumulants. The magnitude of the $n$-th order of the azimuthal anisotropy, $v_{n}$ ($n=1,2,3,...$) is denoted as $v_n\{2k\}$ ($k=1,2,3...$) for the Q-cumulant of the order $2k$. 

The Q-cumulants $c_{n}\{2k\}$ are multivariate polynomial functions of the weighted means of $2m$-particle azimuthal angle ($\phi$) correlations~\cite{Bilandzic:2010jr}
\begin{equation}
\label{2mCumul}
\llangle 2m \rrangle = \frac{\displaystyle\sum^{events}_{i}(W_{\langle 2m \rangle})_{i}\langle 2m \rangle_{i}}{\displaystyle\sum^{events}_{i}(W_{\langle 2m \rangle})_{i}}, \; m=1,...,k
\end{equation}
where the $2m$-particle azimuthal correlation is defined as
\begin{eqnarray}
  \label{single-event}
  \left.\begin{aligned}
&\langle 2m \rangle = \langle e^{in(\phi_{1}+...+\phi_{m}-\phi_{m+1}-...-\phi_{2m})} \rangle = \\
& \frac{(M-2m){\displaystyle !\,}}{M{\displaystyle !\,}}{\displaystyle\sum^{M}_{i_{1}\neq ... \neq i_{2m}=1}}e^{in(\phi_{i_{1}}+...+\phi_{i_{m}}-\phi_{i_{m+1}}-...-\phi_{i_{2m}})},
\end{aligned}\right.
\end{eqnarray}
with $M$ denoting the multiplicity, the number of used charged tracks of the event, and $\sum$ goes over all particle indices that have to be different. The inner brackets $\langle...\rangle$ denote averaging over all unique $2m$-particle multiplets of interest, while the outer brackets $\llangle...\rrangle$ denotes averaging over all events from the given class. $W_{\langle 2m \rangle}$ are the event weights, which are used to minimize the effect of multiplicity variations in the event sample on the estimates of $2m$-particle correlations~\cite{Bilandzic:2010jr}. Instead of using unit or multiplicity itself as a weight, Bilandzic showed~\cite{Bilandzic:2012wva} an obvious advantage of choosing the number of distinct $2m$ particle combinations that one can form for an event with multiplicity $M$ as the corresponding weight: 
\begin{equation}
\label{weight}
W_{\langle 2m \rangle} = \displaystyle\prod^{2m-1}_{j=0}(M-j) = \frac{M!}{(M-2m)!}.
\end{equation}

As suggested in Ref.~\cite{Bilandzic:2010jr}, instead of calculating all possible particle multiplets, one can express cumulants through the corresponding flow vector $Q_{n}$ defined as
\begin{equation}
\label{Q_n}
Q_{n} = {\displaystyle\sum^{M}_{i=1}}e^{in\phi_{i}}.
\end{equation}
The calculations of the flow vector $Q_{n}$ take little amount of time in comparison to the time needed for the calculations of the $2m$-particle azimuthal correlations $\langle 2m \rangle$. Ref.~\cite{Bilandzic:2010jr} gives relations between the flow vectors $Q_{n}$ and the $2m$-particle azimuthal correlations $\langle 2m \rangle$. The corresponding Q-cumulants $c_{n}\{2k\}$ (up to the 8-th order) are given as~\cite{Bilandzic:2010jr}:
\begin{eqnarray}
  \label{Cumul}
    \left.\begin{aligned}
      c_{n}\{2\} &= \llangle 2 \rrangle, \\
      c_{n}\{4\} &= \llangle 4 \rrangle - 2 \llangle 2 \rrangle^{2}, \\
      c_{n}\{6\} &= \llangle 6 \rrangle - 9 \llangle 4 \rrangle \llangle 2 \rrangle + 12 \llangle 2 \rrangle^{3}, \\
      c_{n}\{8\} &= \llangle 8 \rrangle - 16 \llangle 2 \rrangle \llangle 6 \rrangle - 18 \llangle 4 \rrangle^{2} \\
      &+ 144 \llangle 4 \rrangle \llangle 2 \rrangle^{2} - 144 \llangle 2 \rrangle^{4}.
      \end{aligned}\right.
\end{eqnarray}
Finally, the magnitudes of the azimuthal anisotropies $v_{n}$ are related to the above defined multi-particle Q-cumulants through the following equations
\begin{eqnarray}
  \label{FCoeff}
    \left.\begin{aligned}
      v_{n}\{2\} &= \sqrt{c_{n}\{2\}}, \\
      v_{n}\{4\} &= \sqrt[4]{-c_{n}\{4\}}, \\
      v_{n}\{6\} &= \sqrt[6]{\frac{1}{4}c_{n}\{6\}}, \\
      v_{n}\{8\} &= \sqrt[8]{-\frac{1}{33}c_{n}\{8\}}.
      \end{aligned}\right.
\end{eqnarray}

In order to obtain the statistical uncertainties of the $v_n\{2k\}$ harmonics, one commonly divides experimental data into many sub-sets and determines the statistical dispersion of the results or applies the bootstrapping method~\cite{Efron1,Efron2} using today’s readily available computational power. In the case when one uses a small number of data sub-sets or re-sampling, the estimated statistical uncertainty could be unstable, while the analytic calculation can reveal the real uncertainty. Thus, it is challenging to obtain all the statistical uncertainties of the Q-cumulants by calculating them from the data. The first attempt to calculate the statistical uncertainties within the method of cumulants has been performed in Ref.~\cite{Borghini:2001vi}. As the weighted means of $2m$-particle azimuthal correlations given by Eq.~(\ref{2mCumul}) are not mutually independent, knowledge of their variances and all their mutual covariances is needed. Several approaches of calculating these variances and covariances have been proposed. It will be shown that the best approach must include squared weights in the calculation of both the variances and covariances.

In this paper, we present analytical formulas to calculate statistical uncertainties of the $v_{n}$ magnitudes obtained from the Q-cumulants of different orders. Sect.~\ref{Recurrent} gives a recurrence equation that relates Q-cumulants of an arbitrary order. Sect.~\ref{VarCoVar} gives a short overview of finding variances and covariances of the weighted means. Sect.~\ref{StatUncert} gives formulas for the statistical uncertainties of the $v_n\{2k\}$ harmonics. Sect.~\ref{sec:res} verifies the results using a simple toy model simulation of elliptic flow azimuthal anisotropy. A summary is given in Sect.~\ref{sec:conc}.

\section{Recurrence relation between Q-cumulants of any order}
\label{Recurrent}
The Q-cumulant generating function is a logarithm of the weighted means of $2m$-particle azimuthal correlations generating functions~\cite{Borghini:2001vi}. A systematic procedure for analyzing cumulants of an arbitrary order has been presented in~\cite{DiFrancesco:2016srj}. In Ref.~\cite{Smith:1995}, P.~J. Smith gave a recurrence relation between cumulants and moments. The explicit expression for the $2k$-th order of the Q-cumulant, $c_{n}\{2k\}$, in terms of the weighted means of $2m$-particle azimuthal correlations, $\llangle 2m \rrangle$, can be obtained by using Fa\`a di Bruno's formula for higher derivatives of composite functions:
\begin{eqnarray}
\label{c-general}
\left.\begin{aligned}
c_{n}\{2k\} &= \frac{(k{\displaystyle !\,})^{2}}{(2k{\displaystyle !\,})}{\displaystyle\sum^{k}_{m=1}}(-1)^{m-1}(m-1){\displaystyle !\,}\\
& \cdot B_{2k,m}(0,\mu_{2},0,...,\mu_{2k-m+1}),
\end{aligned}\right.
\end{eqnarray}
where $B_{2k,m}$ are the incomplete (or partial) Bell polynomials with the following central moments:
\begin{equation}
\label{moment}
\mu_{2k} = \frac{(2k{\displaystyle !\,})}{(k{\displaystyle !\,})^{2}}\llangle 2k \rrangle, \; \mu_{2k-1}=0, \;k\in N.
\end{equation}
By using the above two Equations~(\ref{c-general}) and (\ref{moment}) is established the following recurrence equation that relates Q-cumulants of an arbitrary order with the weighted means of $2m$-particle azimuthal correlations:
\begin{equation}
\label{Q-general}
c_{n}\{2k\} = \llangle 2k \rrangle - {\displaystyle\sum^{k-1}_{m=1}} \binom{k}{m} \binom{k-1}{m} \llangle 2m \rrangle c_{n}\{2k-2m\}.
\end{equation}
This provide an efficient way to calculate the Q-cumulants.

\section{The $v_{n}\{2k\}$ statistical uncertainties}
\label{Theory}
\subsection{The variances and covariances}
\label{VarCoVar}
The variance of the ordinary arithmetic mean is easily calculated from the variance of the variable itself:
\begin{equation}
\label{AritVar}
\sigma^{2}_{\bar{x}} = \frac{\sigma^{2}_{x}}{N}= \frac{\displaystyle\sum^{N}_{i=1}[x_{i}-\bar{x}]^{2}}{N(N-1)},
\end{equation}
while the covariance is defined as
\begin{equation}
\label{CoVar}
\sigma_{\bar{x}\bar{y}} = \frac{\displaystyle\sum^{N}_{i=1}[x_{i}-\bar{x}][y_{i}-\bar{y}]}{N(N-1)},
\end{equation}
where $N$ denotes the number of single measurements.

However, if one deals with the weighted means, the problem of finding their variances and covariances is getting much more complicated. Gatz and Smith~\cite{Gatz} in 1995 described this in the following way: ``Although the weighted mean is a very useful concept (in precipitation chemistry), it has one notable drawback: there is no readily derivable, generally applicable, analog of standard error of the mean to express the uncertainty of the (precipitation) weighted mean. A theoretical mathematical-statistical development of a formula for the standard error of the weighted mean would require knowledge of the statistical distributions''.

Several expressions for the standard error of the mean have been proposed and used in the literature. Miller in Ref.~\cite{Miller} suggested the following equation:
\begin{equation}
\label{Miler}
\sigma^{2}_{\bar{x}_{w}} = \frac{\displaystyle\sum^{N}_{i=1} w_{i}[x_{i}-\bar{x}_{w}]^{2}}{N\displaystyle\sum^{N}_{i=1} w_{i}},
\end{equation}
that was also used by Liljestrand and Morgan in Ref.~\cite{Liljestrand} and by Topol et al. in Ref.~\cite{Topol}. Here, the weighted mean, $\bar{x}_{w}$, is defined as: $\bar{x}_{w} = \sum^{N}_{i=1} w_{i}x_{i}/\sum^{N}_{i=1} w_{i}$. Although Eq.~(\ref{Miler}) might be seen as the weighted case of Eq.~(\ref{AritVar}), the proposed expression on the right side of Eq.~(\ref{Miler}) was unaccompanied by any discussion or justification of the assumptions required in their derivation.

Galloway et al.~\cite{Galloway} used a somewhat different equation:
\begin{equation}
\label{Galoway}
\sigma^{2}_{\bar{x}_{w}} = \frac{N\displaystyle\sum^{N}_{i=1} w^{2}_{i}x^{2}_{i} - \Big(\displaystyle\sum^{N}_{i=1} w_{i}x_{i}\Big)^{2}}{(N-1)\Big(\displaystyle\sum^{N}_{i=1} w_{i}\Big)^{2}}.
\end{equation}
The equation for the corresponding covariance is then given as
\begin{equation}
\label{GalowayCov}
\sigma_{\bar{x}_{w},\bar{y}_{w}} = \frac{N\displaystyle\sum^{N}_{i=1} w^{x}_{i}w^{y}_{i}x_{i}y_{i} - \displaystyle\sum^{N}_{i=1} w^{x}_{i}x_{i}\displaystyle\sum^{N}_{i=1} w^{y}_{i}y_{i}}{(N-1)\displaystyle\sum^{N}_{i=1} w^{x}_{i}\displaystyle\sum^{N}_{i=1} w^{y}_{i}}.
\end{equation}

A. Bilandzic~\cite{Bilandzic:2012wva} proposed the following variances and covariances to be used in the calculations of the $v_{n}\{2k\}$ statistical uncertainties
\begin{equation}
\label{BilVar}
\sigma^{2}_{\bar{x}_{w}} = \frac{\displaystyle\sum^{N}_{i=1} w^{2}_{i}}{\Big(\displaystyle\sum^{N}_{i=1} w_{i}\Big)^{2}}\frac{\displaystyle\sum^{N}_{i=1} w_{i}[x_{i}-\bar{x}_{w}]^{2}}{\displaystyle\sum^{N}_{i=1} w_{i}}\frac{1}{1-\frac{\sum^{N}_{i=1} w^{2}_{i}}{(\sum^{N}_{i=1} w_{i})^{2}}},
\end{equation}
and
\begin{eqnarray}
\label{BilCov}
\left.\begin{aligned}
\sigma_{\bar{x}_{w},\bar{y}_{w}} & = B_{F}\frac{\displaystyle\sum^{N}_{i=1} w^{x}_{i}w^{y}_{i}}{\displaystyle\sum^{N}_{i=1} w_{i}^{x}\displaystyle\sum^{N}_{i=1} w_{i}^{y}} \\
&  \cdot \Bigg[\frac{\displaystyle\sum^{N}_{i=1} w^{x}_{i}w^{y}_{i}x_{i}y_{i}}{\displaystyle\sum^{N}_{i=1} w_{i}^{x}w_{i}^{y}} - \frac{\displaystyle\sum^{N}_{i=1} w^{x}_{i}x_{i}}{\displaystyle\sum^{N}_{i=1} w_{i}^{x}}\frac{\displaystyle\sum^{N}_{i=1} w^{y}_{i}y_{i}}{\displaystyle\sum^{N}_{i=1} w_{i}^{y}}\Bigg],
\end{aligned}\right.
\end{eqnarray}
where $B_{F} = 1/(1-\frac{\sum^{N}_{i=1} w^{x}_{i}w^{y}_{i}}{\sum^{N}_{i=1} w_{i}^{x}\sum^{N}_{i=1} w_{i}^{y}})$.

Cochran  in his book~\cite{Cochran} has analyzed a problem of finding the variance of the ratio of two random variables. By recognizing the usefulness of that result, Endlich et al.~\cite{Endlich} expressed the variance of the weighted mean as an approximate ratio variance given by Cochran~\cite{Cochran}:
\begin{equation}
\label{Cohran}
\sigma^{2}_{\bar{x}_{w}} = \frac{N}{N-1}\frac{\displaystyle\sum^{N}_{i=1} w^{2}_{i}[x_{i} - \bar{x}_{w}]^{2}}{\Big(\displaystyle\sum^{N}_{i=1} w_{i}\Big)^{2}},
\end{equation}
Gatz and Smith~\cite{Gatz} have shown by bootstrapping methods~\cite{Efron1,Efron2} that the expression on the right side of Eq.~(\ref{Cohran}) gives an accurate estimation for the variance of the weighted mean. Cochran~\cite{Cochran} has also given an approximate expression of the covariance between two ratios. By following the example of Endlich et al.~\cite{Endlich} we are proposing to express the covariance between weighted means as:
\begin{equation}
\label{CohranWMeanCov}
\sigma_{\bar{x}_{w},\bar{y}_{w}} = \frac{N}{N-1}\frac{\displaystyle\sum^{N}_{i=1} w^{x}_{i}w^{y}_{i}[x_{i} - \bar{x}_{w}][y_{i} - \bar{y}_{w}]}{\displaystyle\sum^{N}_{i=1} w_{i}^{x}\displaystyle\sum^{N}_{i=1} w_{i}^{y}},
\end{equation}
where $w^{x}$ and $w^{y}$ are weights of the corresponding variables. This expression is probably not unknown in the literature. Namely, in a recently published article~\cite{Delchambre} the author used formulas for the weighted variance and covariance of a discrete variable. Dealing with the uncorrelated observations with all equal variances there is a known relation between the variance of a weighted variable and the variance of the weighted means:
\begin{equation}
\label{Delch}
\sigma^{2}_{\bar{x}} = \sigma^{2}_{x}\frac{\displaystyle\sum^{N}_{i=1} w^{2}_{i}}{\Big(\displaystyle\sum^{N}_{i=1} w_{i}\Big)^{2}},
\end{equation}
and the corresponding relation between the covariances:
\begin{equation}
\label{DelchCov}
\sigma_{\bar{x},\bar{y}} = \sigma^{2}_{x,y}\frac{\displaystyle\sum^{N}_{i=1} w^{x}_{i}w^{y}_{i}}{\displaystyle\sum^{N}_{i=1} w_{i}^{x}\displaystyle\sum^{N}_{i=1} w_{i}^{y}}.
\end{equation}
It is easy to recognize the connection between the expressions on the right side of Eq.~(\ref{Cohran}) and (\ref{CohranWMeanCov}) and the corresponding expressions in Eq.~(\ref{Delch}) and (\ref{DelchCov}) which are used in~\cite{Delchambre}. The Bessel's correction factors, $\frac{N}{N-1}$ in Eq.~(\ref{Cohran}) and Eq.~(\ref{CohranWMeanCov}), and $\frac{1}{1-\frac{\sum^{N}_{i=1} w^{2}_{i}}{(\sum^{N}_{i=1} w_{i})^{2}}}$ in Eq.~(\ref{BilVar}) and $\frac{1}{1-\frac{\sum^{N}_{i=1} w^{x}_{i}w^{y}_{i}}{\sum^{N}_{i=1} w_{i}^{x}\sum^{N}_{i=1} w_{i}^{y}}}$ in Eq.~(\ref{BilCov}) are very close to one.

\subsection{The $v_{n}\{2k\}$ statistical uncertainties}
\label{StatUncert}
Propagating back to the $\llangle 2m \rrangle$ azimuthal correlations, the statistical uncertainties of the $v_{n}\{2k\}$ expressed by Eq.~(\ref{FCoeff}) are given as
\begin{eqnarray}
  \label{DeltaCumul}
    \left.\begin{aligned}
      &s^{2}[v_{n}\{2\}]\cdot 4(v_{n}\{2\})^{2} = \sigma^{2}_{\scaleto{\llangle 2 \rrangle}{8pt}} \\
      &s^{2}[v_{n}\{4\}]\cdot 16(v_{n}\{4\})^{6} = 16 \llangle 2 \rrangle^{2} \sigma^{2}_{\scaleto{\llangle 2 \rrangle}{8pt}} - 8 \llangle 2 \rrangle \sigma_{\scaleto{\llangle 2 \rrangle, \llangle 4 \rrangle}{8pt}} \\
      &+ \sigma^{2}_{\scaleto{\llangle 4 \rrangle}{8pt}} \\
      &s^{2}[v_{n}\{6\}]\cdot 576(v_{n}\{2\})^{10} = A^{2} \sigma^{2}_{\scaleto{\llangle 2 \rrangle}{8pt}} +2AB\sigma_{\scaleto{\llangle 2 \rrangle, \llangle 4 \rrangle}{8pt}} \\
      &+ B^{2}\sigma^{2}_{\scaleto{\llangle 4 \rrangle}{8pt}} + 2A\sigma_{\scaleto{\llangle 2 \rrangle, \llangle 6 \rrangle}{8pt}} + 2B\sigma_{\scaleto{\llangle 4 \rrangle, \llangle 6 \rrangle}{8pt}} + \sigma^{2}_{\scaleto{\llangle 6 \rrangle}{8pt}} \\
      &s^{2}[v_{n}\{8\}]\cdot 264^{2}(v_{n}\{2\})^{14} = C^{2}\sigma^{2}_{\scaleto{\llangle 2 \rrangle}{8pt}} + D^{2}\sigma^{2}_{\scaleto{\llangle 4 \rrangle}{8pt}} + E^{2}\sigma^{2}_{\scaleto{\llangle 6 \rrangle}{8pt}} \\
      & +\sigma^{2}_{\scaleto{\llangle 8 \rrangle}{8pt}} + 2CD\sigma_{\scaleto{\llangle 2 \rrangle, \llangle 4 \rrangle}{8pt}} + 2CE\sigma_{\scaleto{\llangle 2 \rrangle, \llangle 6 \rrangle}{8pt}} + 2C\sigma_{\scaleto{\llangle 2 \rrangle, \llangle 8 \rrangle}{8pt}} \\
      & + 2DE\sigma_{\scaleto{\llangle 4 \rrangle, \llangle 6 \rrangle}{8pt}} + 2D\sigma_{\scaleto{\llangle 4 \rrangle, \llangle 8 \rrangle}{8pt}} + 2E\sigma_{\scaleto{\llangle 6 \rrangle, \llangle 8 \rrangle}{8pt}},
      \end{aligned}\right.
\end{eqnarray}
where $A$, $B$, $C$, $D$ and $E$ are defined as
\begin{eqnarray}
  \label{ABC}
  \left.\begin{aligned}
      A &= 36\llangle 2 \rrangle^{2} - 9\llangle 4 \rrangle \\
      B &= - 9\llangle 2 \rrangle \\
      C &= 288\llangle 2 \rrangle \llangle 4 \rrangle - 576\llangle 2 \rrangle^{3} - 16\llangle 6 \rrangle\\
      D &= 144\llangle 2 \rrangle^{2} - 36\llangle 4 \rrangle \\
      E &= -16\llangle 2 \rrangle
      \end{aligned}\right.
\end{eqnarray}
In Eq.~(\ref{DeltaCumul}) and (\ref{ABC}), instead of use a general notation of $\bar{x}$ we used a more common notation of $\llangle ... \rrangle$ to denote weighted means of $2m$-particle azimuthal angle correlations.

A. Bilandzic~\cite{Bilandzic:2012wva} gave formulas for the calculations of the statistical uncertainties that are equal to the above formulas presented by the Equations~(\ref{DeltaCumul}) and (\ref{ABC}).

\section{Results}
\label{sec:res}
Via the equations~(\ref{DeltaCumul}) and (\ref{ABC}) one can calculate the statistical uncertainties of the $v_{n}\{2k\}$ harmonics using different expressions for the variances and covariances presented in subsection~\ref{VarCoVar}. The validity of these equations have been tested by simulating various experimental conditions. Beside a uniform multiplicity distribution, a Gauss distribution has been checked. The $(1+2v_{2}cos(2\phi))$ distribution is used to generate particle azimuthal angle. Also, the $v_{2}$ magnitude has been varied according to both Bessel-Gaussian (BG)
\begin{equation}
\label{BG}
f(v_{2}) = \frac{v_{2}}{\sigma_{v}^{2}} e^{-\frac{v_{2}^{2}+\mu^{2}}{2\sigma_{v}^{2}}} I_{0}(\frac{v_{2}\mu}{\sigma_{v}^{2}}),
\end{equation}
where  the flow fluctuation $\sigma_{v}$ is taken to be $3\mu/4.4$, and Gaussian (normal) distribution
\begin{equation}
\label{Gauss}
f(v_{2}) = \frac{1}{\sigma_{v}\sqrt{2\pi}} e^{-\frac{1}{2}(\frac{v_{2}-\mu}{\sigma_{v}})^{2}},
\end{equation}
where the $\sigma_{v}$ is taken to be $1\mu/4.4$. In order to show robustness of the proposed method for the analytic calculations of the statistical uncertainties of the $v_{n}\{2k\}$ magnitudes we performed analyses in which we used small fixed multiplicities ($50$ and $150$), or fixed $v_{2}$ magnitude, or inclusion of the $v_{1}$ harmonic into the simulation of the data. Additionally, the analysis has been repeated several times with the following mean $\mu$ (mean $v_{2}$) values: 5\% and 15\% in the case of the BG distribution, and with 0\%, 2\%, 5\%, 10\%, 20\% and 40\% in the case of the Gaussian distribution. In all of these analyses, qualitatively the same conclusion is reached concerning the consistencies between the calculated statistical uncertainties and those estimated by the data sub-sets and the bootstrapping method. For the sake of brevity, in the analysis presented in this paper, uniform multiplicity distribution and $v_{2}$ magnitude of 5\% varied by BG distribution is used.
\begin{figure}[h!]
  \includegraphics[width=0.5\textwidth]{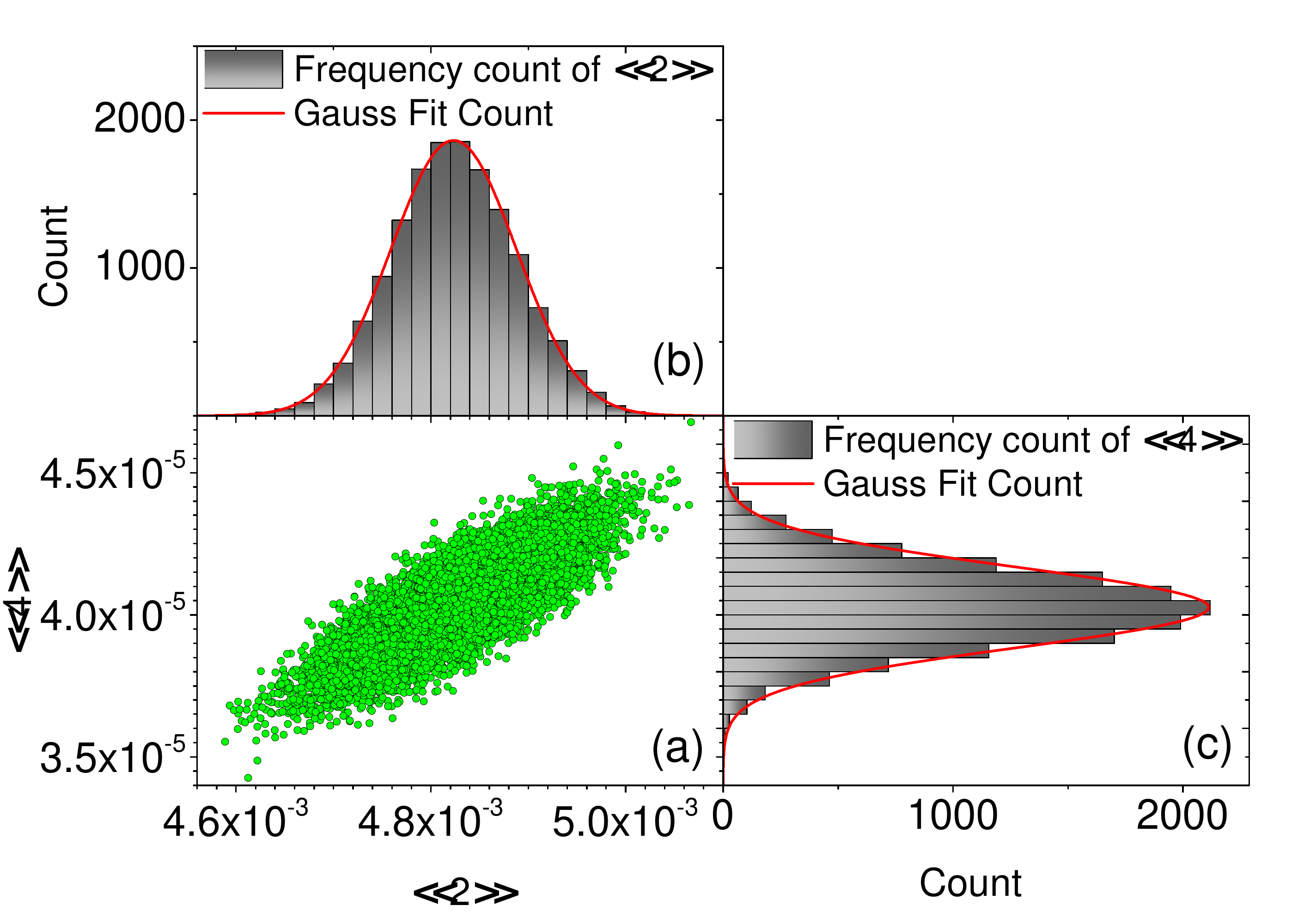}
  \caption{\label{fig:1} (Color online) Two- and four-particle azimuthal correlations obtained by 15000 simulations of 10000 events each with elliptic flow $v_{2}$ magnitude that varies according to the Bessel-Gaussian distribution from event to event, with the mean $v_{2}$ magnitude of 5\% and $\sigma_{v} = 3\mu/4.4$.}
\end{figure}

Presented in Fig.~\ref{fig:1} are the results of calculations of the two- and four-particle azimuthal correlations, $\llangle 2 \rrangle$ and $\llangle 4 \rrangle$, obtained by 15000 simulations of 10000 events each with elliptic flow $v_{2}$ magnitude that varies according to the BG distribution with $\mu = 0.05$ and $\sigma = 3\mu/4.4$ from event to event. The multiplicity of the events has been varied uniformly from 300 to 900. This is chosen in order to roughly simulate experimental conditions in lower energies nucleus-nucleus collisions which will be experimentally available at the NICA collider. At higher collision energies, due to the greater multiplicity $M$ and greater $v_{n}$ magnitudes, the feasibility of the Q-cumulant method is anyhow better. When multiplicity becomes small enough ($M < 300$) the feasibility of the Q-cumulant method deteriorates. The deterioration becomes worse with further decrease of the multiplicity. In addition, the statistical uncertainties increase enormously.

By repeating the simulations under the same experimental conditions one gets a population of the simulated values of those $2m$-particle azimuthal correlations that serves one to compare the dispersions of the obtained distributions with the dispersions (variances) predicted by Equations given in subsection~\ref{VarCoVar}. Each pair of the simulated values $\{\llangle 2 \rrangle, \llangle 4 \rrangle\}$ is presented by a green filled circle in the lower left panel of the Fig.~\ref{fig:1}. We have applied a two-dimensional (2D) binning in that panel and built up the frequency count distributions presented by corresponding column graphs (upper left and lower right panels). The normality of the distributions of both of the populations, $\llangle 2 \rrangle$ and $\llangle 4 \rrangle$, is checked by Anderson-Darling test. It is found that only the population of $\llangle 2 \rrangle$  have passed it at the confidence level of 5\%. Both populations have passed the Kolmogorov-Smirnov normality test at the same confidence level and we fitted both frequency distributions by Gauss functions presented by red color lines in Fig.~\ref{fig:1}. Distributions $\llangle 2 \rrangle$ and $\llangle 4 \rrangle$ yield the $v_{2}$ results with the corresponding statistical uncertainties as 0.069456$\pm$0.000004 and 0.04994$\pm$0.00002 respectively.

\begin{figure}[h!]
  \includegraphics[width=0.5\textwidth]{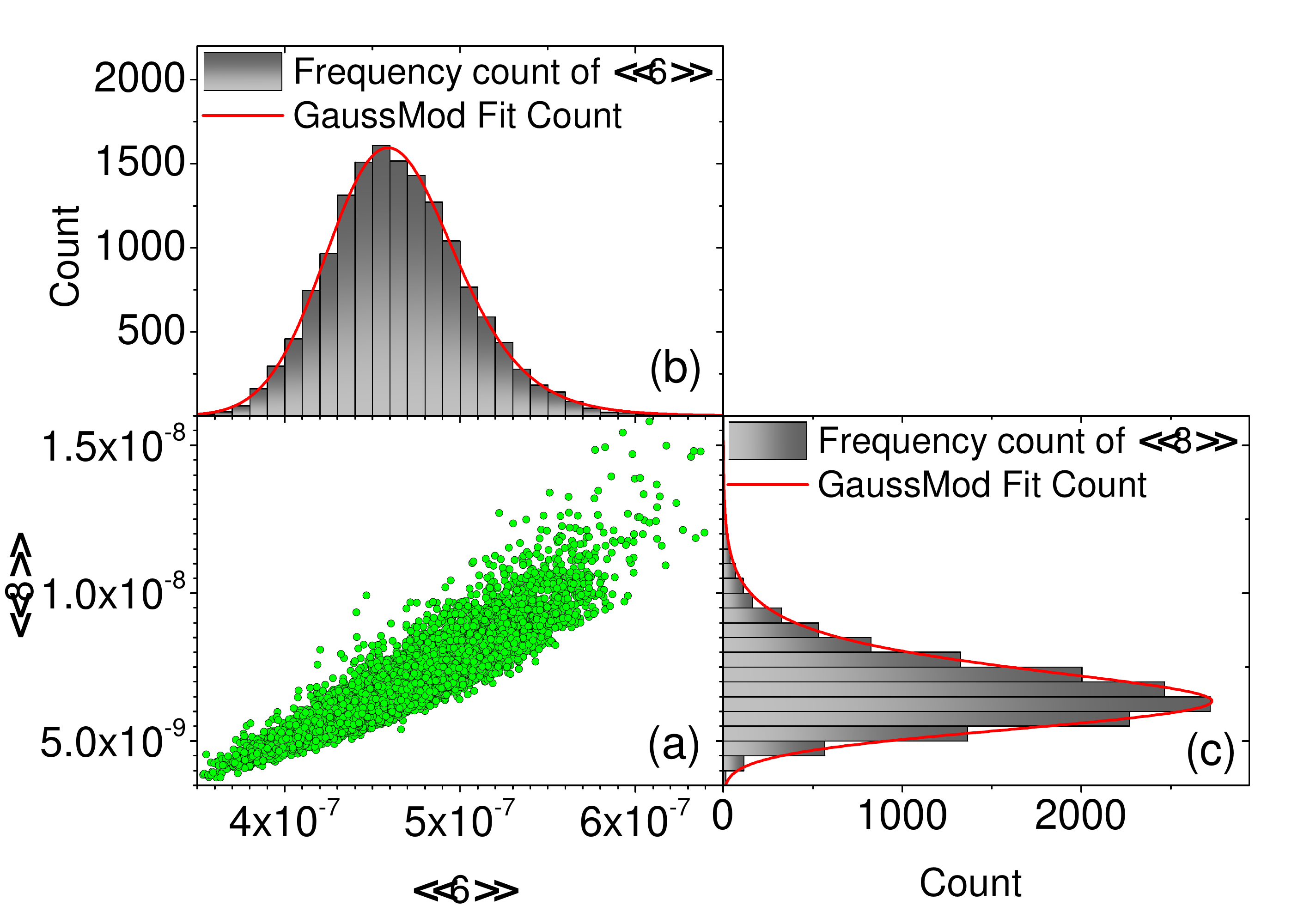}
  \caption{\label{fig:2} (Color online) Six- and eight-particle azimuthal correlations obtained by 15000 simulations of 10000 events each with elliptic flow $v_{2}$ magnitude that varies according to the Bessel-Gaussian distribution from event to event, with the mean $v_{2}$ magnitude of 5\% and $\sigma_{v} = 3\mu/4.4$.}
\end{figure}
The similar analysis is presented in Fig.~\ref{fig:2} for the 6- and 8-particle azimuthal correlations. Neither of the two distributions could pass any normality test at the level of confidence of 5\%. Both distributions have vivid right tailings and they are fitted by the exponentially modified Gaussians. The non-normality of the distributions of the higher number particle correlations calls for caution in the interpretation of the confidence intervals; they are certainly not corresponding to the standard ones. The conditions for the Central Limit Theorem are not fulfilled completely. The $\llangle 6 \rrangle$ and $\llangle 8 \rrangle$ distributions yield the $v_{2}$ results as 0.04980$\pm$0.00002 and 0.04980$\pm$0.00002 respectively.

\begin{figure}[h!]
  \includegraphics[width=0.5\textwidth]{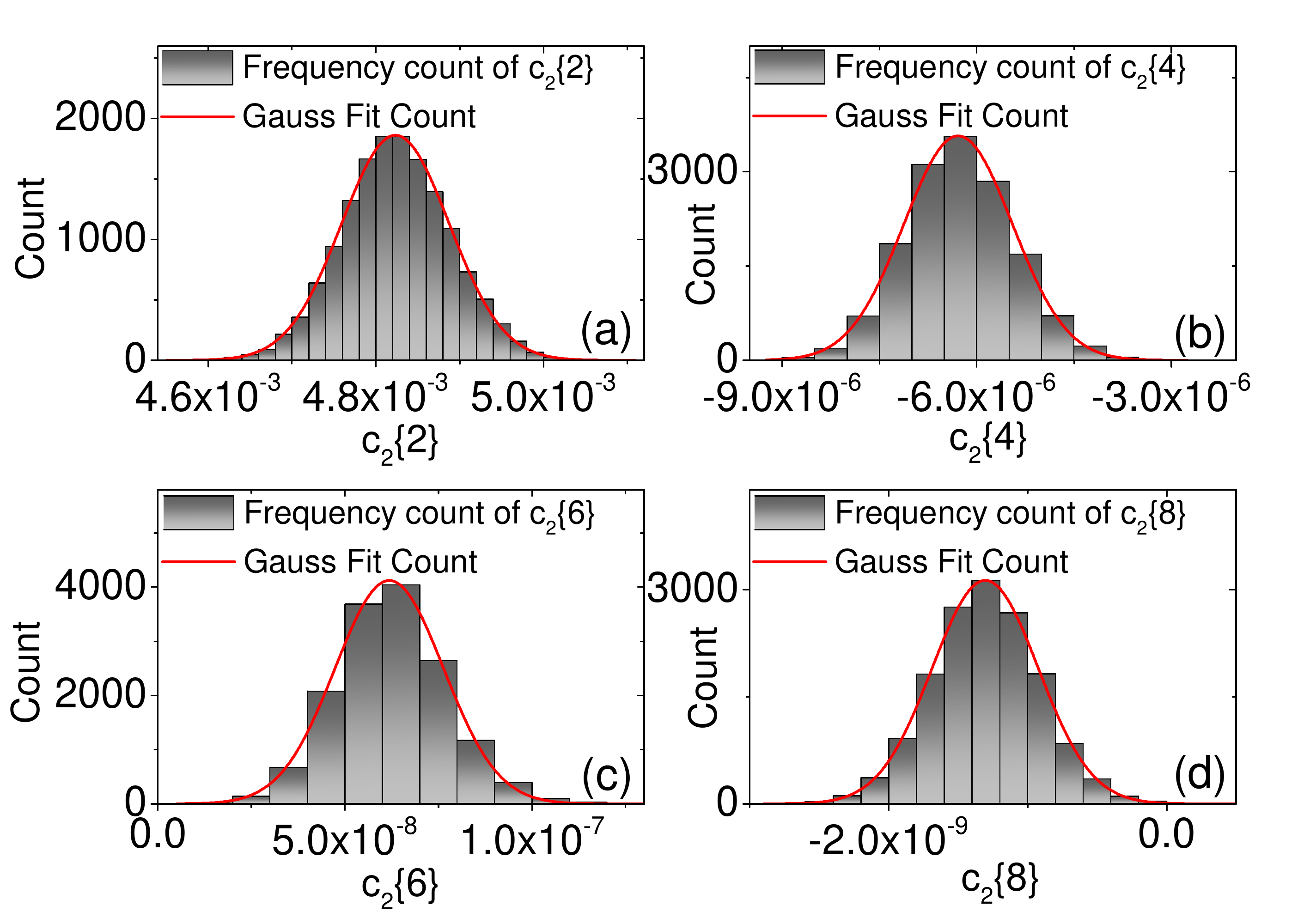}
  \caption{\label{fig:3} (Color online) Frequency counts of the Q-cumulants  $c_{2}\{2k\}$, $k = 1, ... 4$. The red curves represent corresponding Gaussians.}
\end{figure}
Figure~\ref{fig:3} displays the distributions of the Q-cumulants $c_{2}\{2k\}$, $k = 1, ... 4$ values reconstructed using Eq.~(\ref{Cumul}). In contrast to the pronounced skewness of the $\llangle 2m \rrangle$ distributions, that is clearly visible especially for the $m=3,4$ (see Fig.~\ref{fig:2}), in the distributions of the Q-cumulants $c_{n}\{2k\}$ it is not the case.

\begin{figure}
\includegraphics[width=0.5\textwidth]{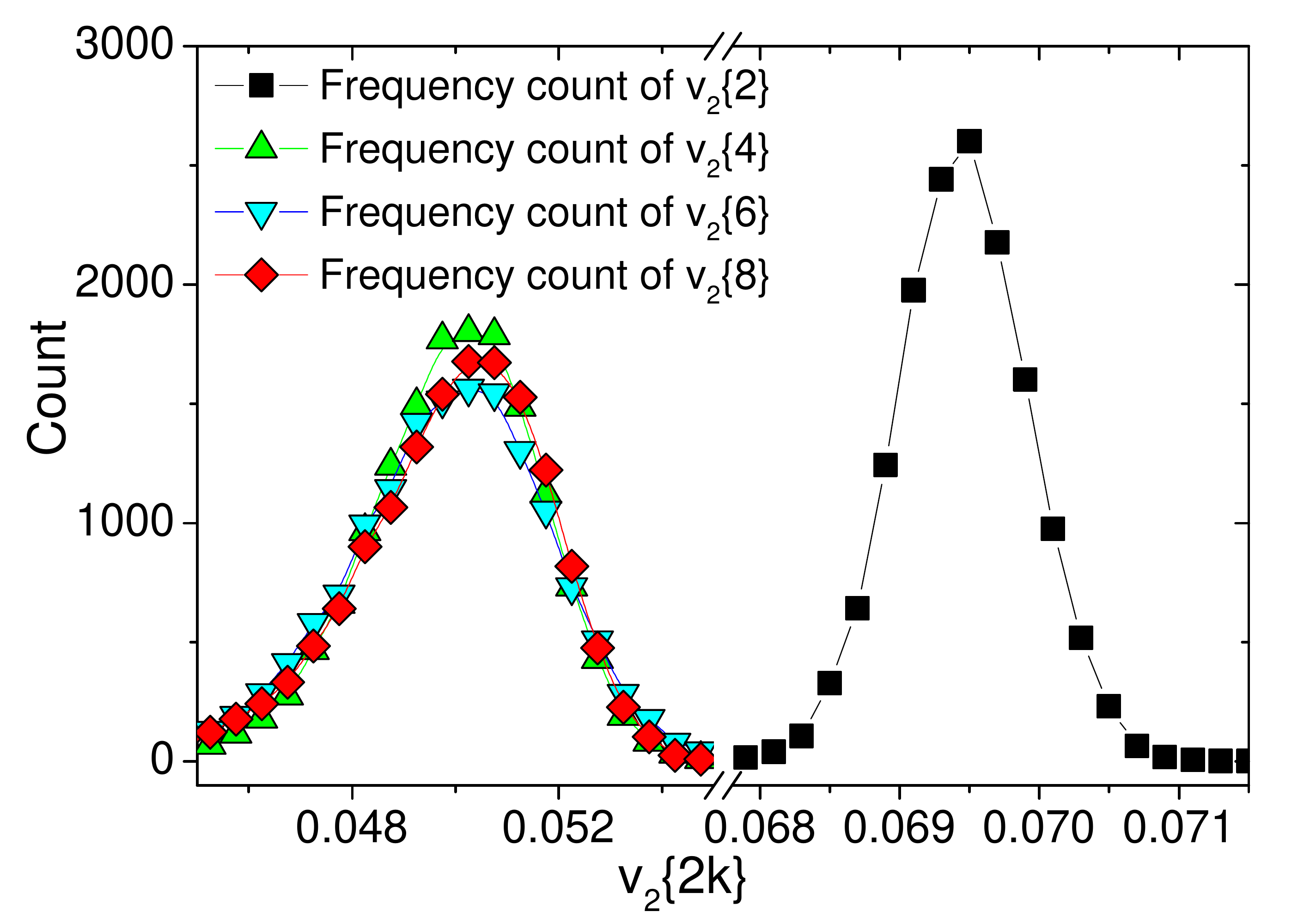}
\caption{\label{fig:4} (Color online) The $v_{2}\{2k\}$ values reconstructed from the corresponding $c_{2}\{2k\}$ Q-cumulants.}
\end{figure}

Shown in Fig.~\ref{fig:4} are the $v_{2}\{2k\}$ values reconstructed from the corresponding $c_{2}\{2k\}$ Q-cumulants. A clear separation between the $v_{2}\{2\}$ from one side, and the $v_{2}\{2k\}$ ($k=2,3,4$) on the other side is visible. When the non-flow effects are negligible, by using the Taylor expansion up to the first order, it has been proven in Ref.~\cite{Ollitrault:2009ie,Bilandzic:2012wva} that due to statistical flow fluctuations the $v_{2}\{2\}$ and higher cumulant order $v_{2}\{2k\}$ are connected via
\begin{equation}
\label{fluct}
v^{2}_{2}\{2\} \approx v^{2}_{2}\{2k\} + 2\sigma^{2}_{v} , \;\; k=2,3,4
\end{equation}
where $\sigma_{v}$ is the flow fluctuations. The obtained results shown in Fig.~\ref{fig:4} nicely reproduce the input $\sigma_{v}$ of $3\cdot 0.05/4.4$.

\begin{figure}[h!]
\includegraphics[width=0.5\textwidth]{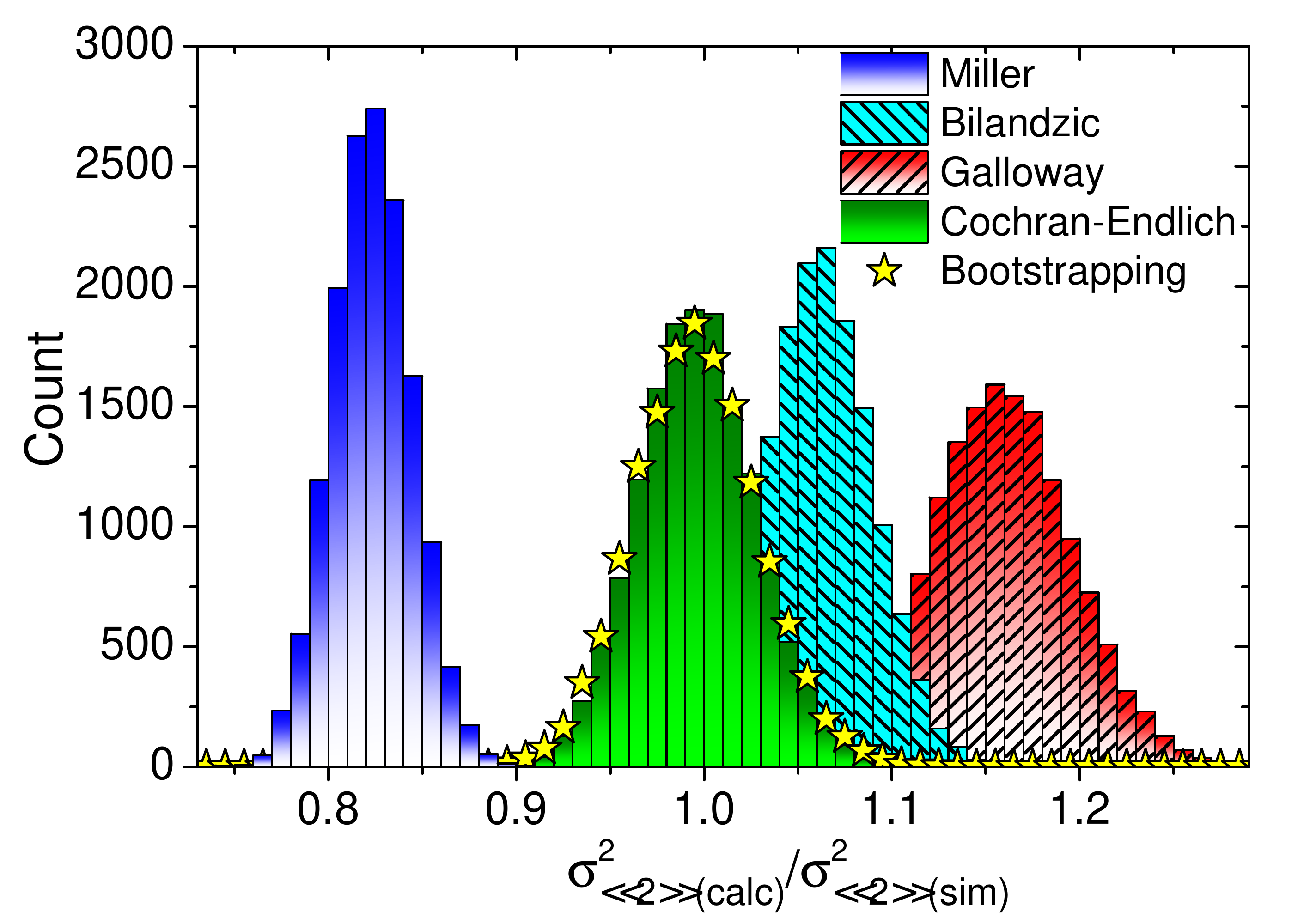}
\caption{\label{fig:5} (Color online) Distributions of the variances of the azimuthal correlations $\llangle 2 \rrangle$ calculated using different expressions presented in subsection~\ref{VarCoVar} over the variances obtained by simulations. Shown with yellow stars is distribution of the corresponding ratio of the bootstrapping results over the results obtained by simulations.}
\end{figure}

As en example, Fig.~\ref{fig:5} shows the distributions of the variances of the azimuthal correlations $\llangle 2 \rrangle$ calculated by using different expressions presented in subsection~\ref{VarCoVar} divided by the variances obtained by the data sub-sets method. The calculations obtained using the right sides of the equations given in subsection~\ref{VarCoVar} are shortly called 'calculated' and marked as 'calc', while those obtained from dispersions of the results from data sub-sets are called 'simulated' and marked as 'sim'. Additionally, Fig.~\ref{fig:5} also shows the distribution of the corresponding ratio of the bootstrapping results over the results obtained by simulations. As expected, the bootstrap results are in an excellent agreement with the simulation results. The same is valid for the results analytically obtained using Eq.~(\ref{Cohran}), too. The variances predicted by Eq.~(\ref{Miler}) are smaller, while those obtained by Eq.(\ref{Galoway}) and Eq.~(\ref{BilVar}) are greater with respect to the simulation results.

\begin{figure}[h!]
\includegraphics[width=0.5\textwidth]{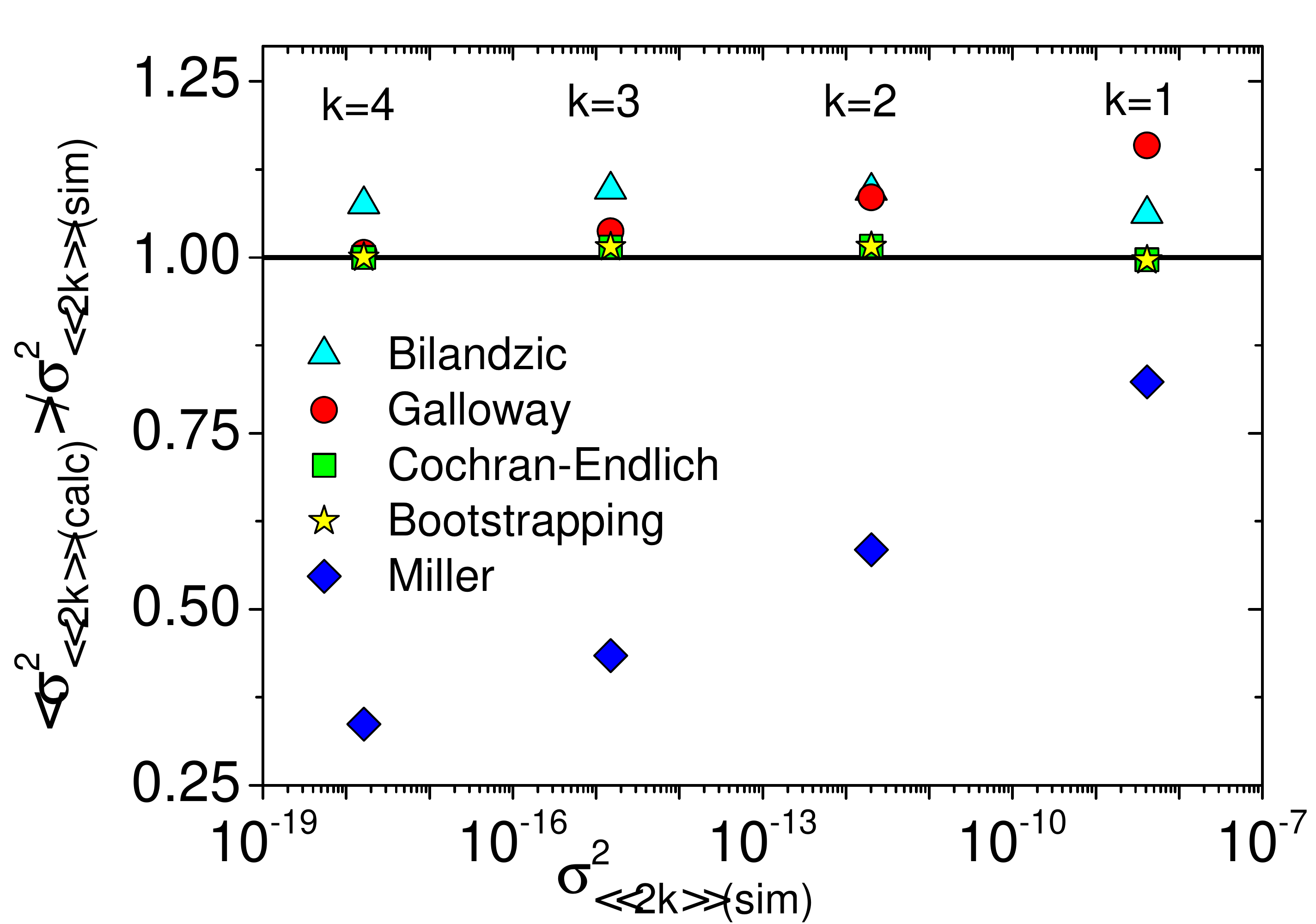}
\caption{\label{fig:6} (Color online) The mean values of the variances of the azimuthal correlations $\llangle 2k \rrangle$, $k = 1, ..., 4$ calculated using different expressions presented in subsection~\ref{VarCoVar} over the mean values of the variances obtained by simulations. Shown with yellow stars are the corresponding mean values from the bootstrapping method.}
\end{figure}

Presented in Fig.~\ref{fig:6} are the ratios of the mean values of the variances of the azimuthal correlations $\llangle 2k \rrangle$, $k = 1, ..., 4$ calculated by using different expressions presented in subsection~\ref{VarCoVar}, over mean variances obtained from simulations. The same ratio is calculated for the results from the bootstrapping method, too. The results obtained by Eq.~(\ref{Miler}) shows great deviations from the simulated variances. The deviations become larger with an increase of $k$. Eq.~(\ref{Galoway}) gives a fair estimation of the variances for the higher orders of the azimuthal correlation, while it starts to deviate for the lower orders $k = 1,2$. Also, results obtained by Eq.~(\ref{BilVar}) deviate for all order of $k$. However, the results obtained by Eq.~(\ref{Cohran}) are in an extraordinary accordance with the simulated results for all orders of $k$. The same is true for the results obtained using the bootstrapping method.

\begin{figure}[h!]
  \includegraphics[width=0.5\textwidth]{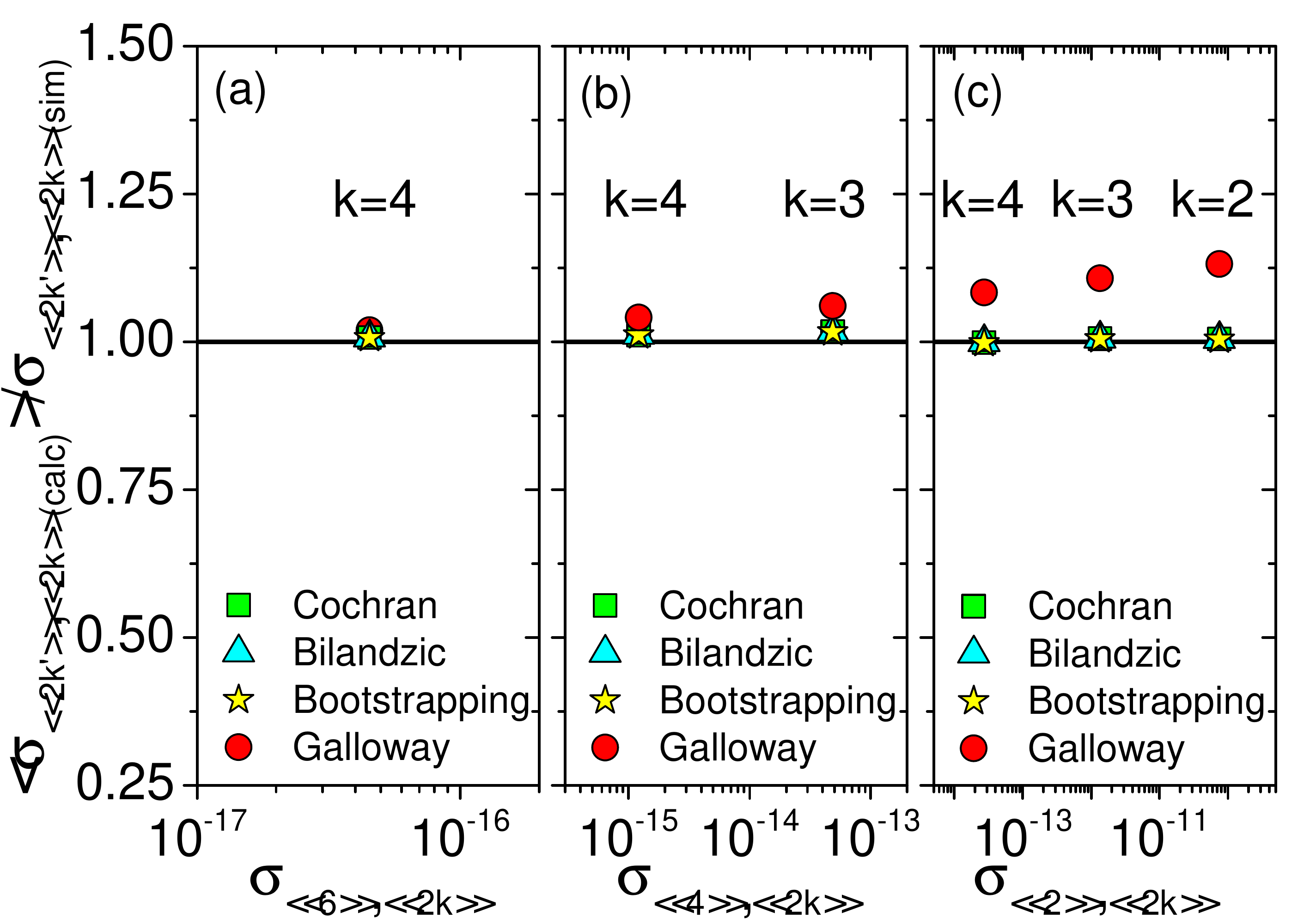}
  \caption{\label{fig:7} (Color online) The ratio of the mutual covariances between azimuthal correlations $\llangle 2k \rrangle$, $k = 1, ..., 4$ calculated using different expressions presented in subsection~\ref{VarCoVar} over the covariances obtained by simulations. With yellow stars are shown the corresponding mean values from the bootstrapping method.}
\end{figure}

Figure~\ref{fig:7} shows the ratio of the mutual covariances of the azimuthal correlations $\llangle 2k \rrangle$, $k = 1, ..., 4$ calculated using different expressions presented in subsection~\ref{VarCoVar} over the covariances obtained by simulations. Additionally, the corresponding mean covariance values from the bootstrapping method are also presented. As there is no an analogous covariance equation for the variance in Eq.~(\ref{Miler}), so these values are omitted in Fig.~\ref{fig:7}. Similarly to Fig.~\ref{fig:6}, Eq.~(\ref{GalowayCov}) for the covariances that is associated to the variance expressed by Eq.~(\ref{Galoway}) gives a fair estimation of the covariances for the higher orders of the azimuthal correlation, $k$, while deviates for the lowest order. In contrast to the results for the variances, the corresponding results for the covariances obtained by using Eq.~(\ref{BilCov}) are in an excellent agreement with the results obtained from the simulations. The reason is that naturally, a product of the weights had to be introduced into the right side of Eq.~(\ref{BilCov}). Again, the same as in the case of the variances, the results for covariances obtained by using Eq.(\ref{CohranWMeanCov}) are in extraordinary accordance with the simulated results in the entire checking region. The same is valid for the results obtained using the bootstrapping method, too.

\begin{figure}[h!]
  \includegraphics[width=0.5\textwidth]{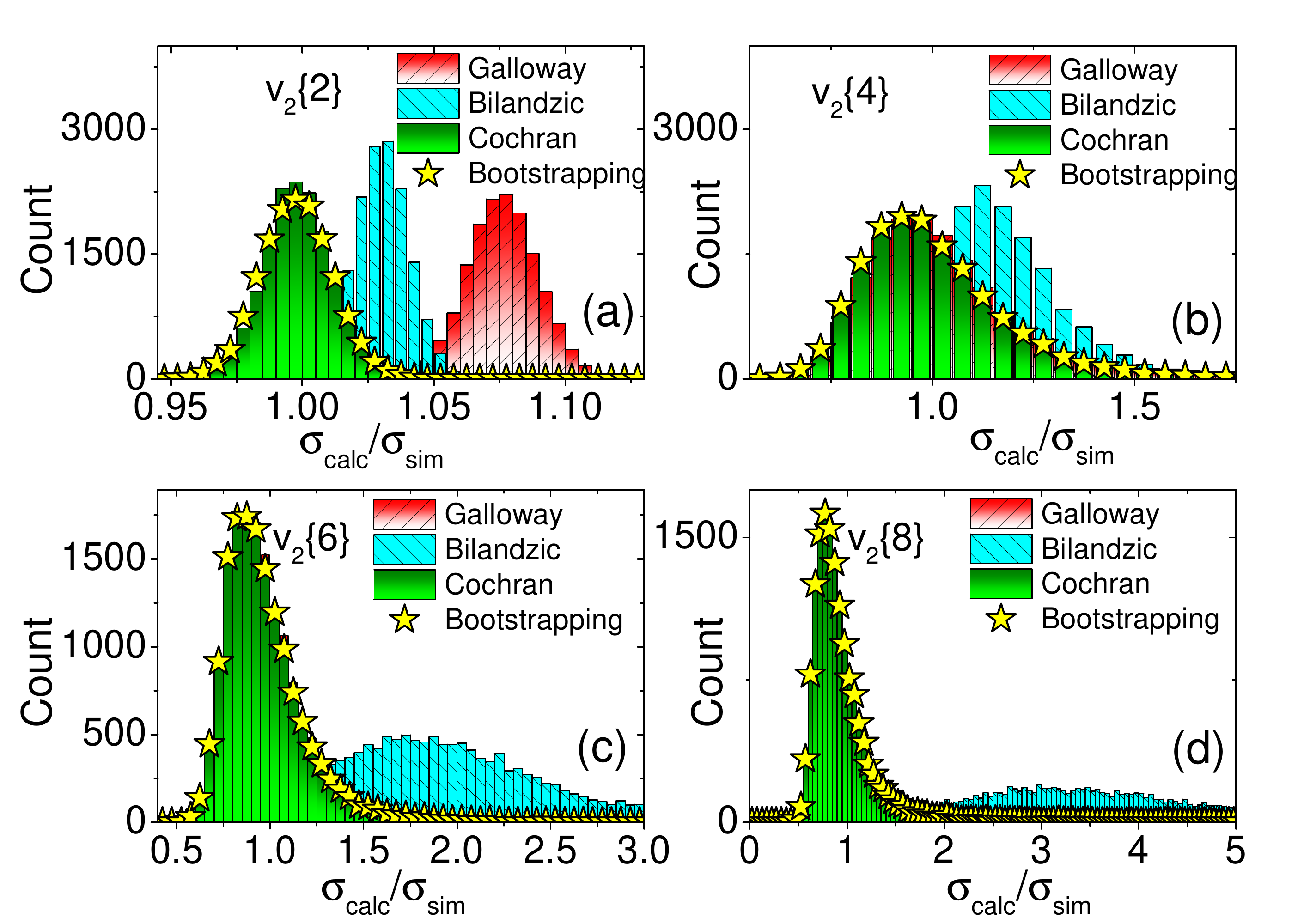}
  \caption{\label{fig:8} (Color online) The distributions of the ratio of the statistical uncertainties of the $v_{2}\{2k\}$, $k = 1, ..., 4$ calculated by Eq.~(\ref{DeltaCumul}) using different expressions presented in subsection~\ref{VarCoVar} over the statistical uncertainties of the $v_{2}\{2k\}$ obtained by simulations. Shown with yellow stars are the corresponding distributions obtained from the bootstrapping method.}
\end{figure}

Finally, Fig.~\ref{fig:8} shows distributions of the ratio of the statistical uncertainties of the $v_{2}\{2k\}$, $k = 1, ..., 4$ calculated by Eq.~(\ref{DeltaCumul}) using different expressions presented in subsection~\ref{VarCoVar} over the statistical uncertainties of the $v_{2}\{2k\}$ obtained by simulations. Additionally, the corresponding results obtained from the bootstrapping method are shown too. As expected, they are in an excellent agreement with the simulation results. The same conclusion is valid when one use Eq.~(\ref{Cohran}) and Eq.~(\ref{CohranWMeanCov}) to calculate the $v_{2}\{2k\}$ statistical uncertainties. The $v_{2}\{2k\}$ statistical uncertainties obtained using Eq.~(\ref{Galoway}) and Eq.~(\ref{GalowayCov}) for variances and covariances respectively are greater than they should be, but the deviation becomes somewhat smaller going to higher cumulant order. In the case of using Eq.~(\ref{BilVar}) and Eq.~(\ref{BilCov}) the deviation is big and becomes greater with an increase of the cumulant order. For the $k=4$, the $v_{2}\{2k\}$ statistical uncertainty becomes nearly four times larger than it should be. The source of the deviation is entirely in the use of Eq.~(\ref{BilVar}) to calculate the variance where the weights are introduced linearly, instead of quadratically. Thus, although the weights used in this paper are the same as those introduced in~\cite{Bilandzic:2010jr,Bilandzic:2012wva}, the way how they are implemented in the formula for the variance makes a significant difference.

\begin{figure}[h!]
  \includegraphics[width=0.5\textwidth]{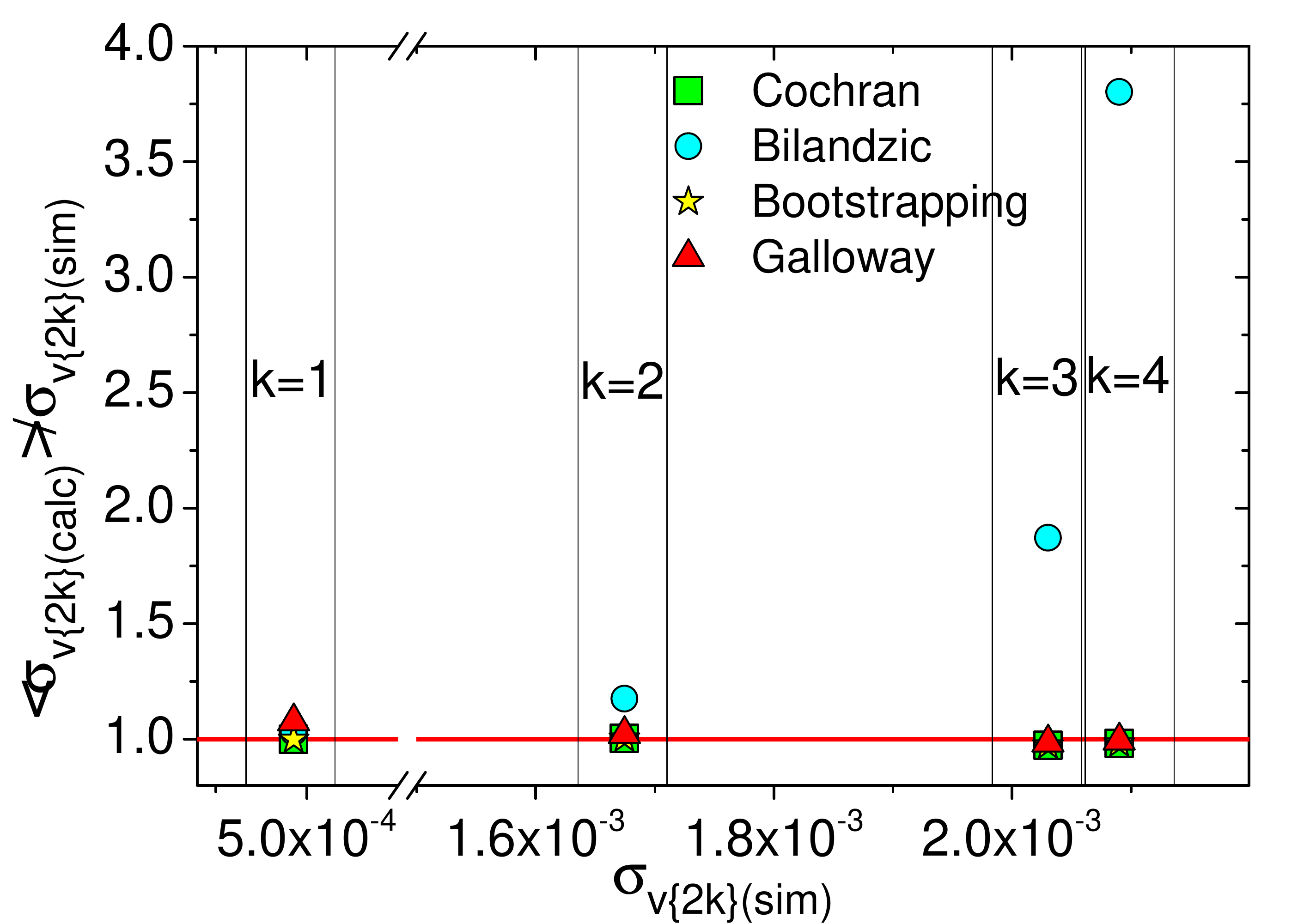}
  \caption{\label{fig:9} (Color online) The ratio of the mean statistical uncertainties of the $v_{2}\{2k\}$, $k = 1, ..., 4$ calculated by Eq.~(\ref{DeltaCumul}) using different expressions presented in subsection~\ref{VarCoVar} over the statistical uncertainties of the $v_{2}\{2k\}$ obtained by simulations. Shown with yellow stars are the corresponding results obtained from the bootstrapping method.}
\end{figure}

Fig.~\ref{fig:9} summarize the results by plotting the mean values of the corresponding distributions shown in Fig.~\ref{fig:8}. The results show an excellent agreement with the results obtained by simulations when one use the bootstrapping method or analytical calculation of the $v_{2}\{2k\}$ statistical uncertainties based on use of Cohran's Eq.~(\ref{Cohran}) and Eq.~(\ref{CohranWMeanCov}). The use of Galoway's Eq.~(\ref{Galoway}) and Eq.~(\ref{GalowayCov}) for variances and covariances results in somewhat greater $v_{2}\{2k\}$ statistical uncertainties than they should be, while the use of Eq.~(\ref{BilVar}) produce much larger statistical uncertainties.

\section{Conclusions}
\label{sec:conc}
In this paper we presented analytic expressions for calculating the statistical uncertainties of $v_{n}\{2k\}$ harmonics extracted using the Q-cumulants method. The analysis is performed using a simple toy model which simulates elliptic flow azimuthal anisotropy with magnitudes around 0.05. The estimation of the statistical uncertainties of $v_{n}\{2k\}$ is based on the calculation of the variances and covariances of the $\llangle 2m \rrangle$, $m=1,...,k$ azimuthal anisotropies expressed by different equations given in subsection~\ref{VarCoVar}. When one use Cohran's Eq.~(\ref{Cohran}) and Eq.~(\ref{CohranWMeanCov}), for all orders of $k$, an extraordinary accordance is achieved for both variances and covariances between the calculated and those obtained from the dispersion of the results from many data sub-sets. The same is true for the final $v_{n}\{2k\}$ statistical uncertainties. Additionally, the results obtained by the bootstrapping method gives an excellent agreement with the results from data sub-sets. The proposed way of the analytic calculations of the statistical uncertainties of the $v_{n}\{2k\}$ magnitudes is robust to the change of the multilicity, the $v_{n}$ magnitude and inclusion of the other Fourier harmonics. In addition, a recurrence relation between the Q-cumulants of any order is also presented.

\begin{acknowledgments}
The authors acknowledge the support from Ministry of Education Science and Technological Development, Republic of Serbia, National Natural Science Foundation of China (Grant No.~12035006, 12075085, 12047568) and the U.S.~Department of Energy (Grant No.~de-sc0012910).
\end{acknowledgments}

\end{document}